\documentclass[12pt]{article}
\usepackage{epsfig}
\usepackage{amsfonts}
\usepackage{amsopn}
\usepackage{amsmath}
\usepackage{color}

\newcommand{\eq}{\begin{equation}}
\newcommand{\eqx}{\end{equation}}
\newcommand{\eqn}{\begin{eqnarray}}
\newcommand{\eqnx}{\end{eqnarray}}
\newcommand{\f}[2]{\frac{#1}{#2}}

\newcommand{\lm}{\lambda}

\newcommand{\dl}{\delta}

\newcommand{\eps}{\varepsilon}
\newcommand{\qqqq}{\quad\quad\quad\quad}
\newcommand{\nn}{{\cal N}}
\newcommand{\sinpt}{\sin \f{p}{2}}
\newcommand{\qs}{q_*}
\DeclareMathOperator{\arcsinh}{arcsinh}

\title{A new derivation of L{\"u}scher F-term and fluctuations around the
  giant magnon}

\author{Micha{\l} P. Heller\thanks{e-mail: {\tt
      heller@th.if.uj.edu.pl}}, Romuald A. Janik\thanks{e-mail: {\tt
      ufrjanik@if.uj.edu.pl}}\\ and Tomasz {\L}ukowski\thanks{e-mail:
      {\tt tomaszlukowski@gmail.com}} \\ \\
Institute of Physics\\
Jagellonian University,\\
ul. Reymonta 4, \\
30-059 Krak{\'o}w\\
Poland
}

\begin{document}

\maketitle

\begin{abstract}
In this paper we give a new derivation of the generalized L{\"u}scher
F-term formula from a summation over quadratic fluctuations around a
given soliton. The result is very general providing that S-matrix is diagonal
and is valid for arbitrary dispersion relation. We then
apply this formalism to compute the leading finite size corrections to
the giant magnon dispersion relation coming from quantum fluctuations.
\end{abstract}

\section{Introduction}

The discovery of integrability in the context of the AdS/CFT
correspondence \cite{adscft} which appeared both on the gauge theory
side \cite{Minahan:2002ve,Beisert:2003tq,Beisert:2003yb,kor1,
Dolan:2003uh} and on the string theory side \cite{Bena:2003wd}
gives a hope for finding, in principle, the exact spectrum of the
quantized superstring in $AdS_5 \times S^5$ and equivalently the
spectrum of anomalous dimensions of all operators in $\nn=4$ Super
Yang-Mills theory.

A lot of progress has been done in the setting of infinitely long
strings (strings with large charges/angular momenta) or very long
gauge theory operators. The S-matrix for elementary excitations has
been identified, initially in various subsectors \cite{S,BS} and then
for the full multiplet of elementary excitations \cite{Beisert}. The
remaining overall scalar function -- the so-called dressing factor
\cite{AFS,HL} --
has been finally fixed in \cite{BHL,BES} satisfying constraints of crossing
symmetry \cite{CROSSING}.

Despite the immense progress a lot remains to be understood concerning
the structure of energy levels of strings with finite charges or short
operators. On the gauge theory side the problem was identified with
wrapping interactions \cite{BDS}. On the string side of the duality,
these correspond to virtual corrections coming from particles
propagating around
the string worldsheet cylinder \cite{US}. These corrections lead to
effects which go beyond
the asymptotic Bethe ansatz. Such phenomena have been observed in various
calculations \cite{SN,SNZZ,AFZ,Sem,LSRV} and models
\cite{Hubbard,Fioravanti}. Currently intensive work is
being done both at weak \cite{Zanon,Eden,Mann} and at strong
coupling \cite{MinahanSax}. At strong coupling
the finite size effects come (roughly) in two varieties
\eq
\dl\eps \propto e^{-\f{2\pi J}{\sqrt{\lm} \sinpt}}
\eqx
as for the classical finite size correction to the giant magnon
\cite{AFZ}, and
\eq
\label{e.fmagn}
\dl\eps \propto e^{-\f{2\pi J}{\sqrt{\lm}}}
\eqx
which typically arises from a summation over quadratic fluctuations
around a classical string solution.

In \cite{JL}, generalizations of L{\"u}scher formulas \cite{Luscher} were
derived for finding the leading finite-size correction to the
dispersion relation of elementary excitations. There are two types of
contributions: the $\mu$-term and the $F$-term. The first corresponds
to a particle splitting into two on-shell particles which cross the
cylinder and recombine. It is intimately related to the existence of
bound states. At strong coupling it gives the correction
\eq
\dl \eps^\mu = -\f{\sqrt{\lm}}{\pi} \cdot \f{4}{e^2} \cdot \sin^3 \f{p}{2}
\cdot e^{-\f{2\pi J}{\sqrt{\lm} \sinpt}}
\eqx
which exactly reproduces the classical computation of the leading
finite-size giant magnon dispersion relation of \cite{AFZ}. Other
checks have been done dealing with dyonic magnons \cite{HS}.

The motivation for this work was to explore the role of the $F$-term
for the giant magnon. One expects that summing the energies of small
fluctuations around the giant magnon solution should give a correction
of the type (\ref{e.fmagn}) which should be reproduced by the $F$-term
formula\footnote{The $(-1)^{F_b}$ missing in the original derivation of
  \cite{JL} has been independently observed in
  \cite{Vieira}. Diagrammatically the $(-1)^{F_b}$ arises from a $-1$
  factor due to a fermion loop in the 1PI self-energy. It can be
  compensated by another $-1$
  if the fermions are antiperiodic on the cylinder (in the TBA
  interpretation this corresponds to a computation of the thermal
  partition function) but this does not happen here and the TBA
  interpretation is rather the computation of an index.}
\eq
\label{e.fgen}
\dl \eps^{F}_a = -\int_{-\infty}^\infty \f{dq}{2\pi}  \left(1-
\f{\eps'(p)}{\eps'(\qs)} \right) \cdot e^{-i\qs L} \cdot \sum_b (-1)^{F_b}
\left( S_{ba}^{ba}(\qs,p)-1 \right)
\eqx
Here $q$ is the original euclidean energy which plays the role of
momentum in the space-time interchanged theory, $E=\eps(p)$ is the
dispersion relation and $\qs$ is determined by the Euclidean on-shell condition
\eq
q^2+\eps^2(\qs)=0
\eqx
It will be convenient to change integration variables and rewrite
(\ref{e.fgen}) as
\eq
\label{e.fgen2}
\dl \eps^{F}_a =\int_{-\infty}^{\infty}\frac{\mbox{d}\qs}{2 \pi
  i}\left(\eps'(\qs)-
\eps'(p) \right)\cdot e^{-i \qs L }\cdot \sum_b (-1)^{F_b}
\left( S_{ba}^{ba}(\qs,p)-1 \right)
\eqx
where we used the relation
\eq
q=\pm i\eps(\qs)
\eqx
and we choose the plus sign.

In the course of performing the calculations we found a very close
link of the generalized formula (\ref{e.fgen}) or (\ref{e.fgen2}) with
the summation over
energies of small fluctuations. Indeed we found that the whole
expression (\ref{e.fgen2}) can be exactly reproduced from a summation
over quadratic fluctuations provided one uses {\em exact} scattering
phase-shifts (and not just semiclassical ones).
The result is very general and does not
depend on the form of the dispersion relation but is true only in the case of
diagonal scattering.

The plan of this paper is as follows. In section 2 we will present the setup
of the quadratic fluctuation calculation. In section 3 we will perform
a Poisson resummation over the energies and recover (\ref{e.fgen2}). In
section 4 we will apply the above formalism to find the leading
quantum correction to the giant magnon. We will close the paper with a
discussion.

\section{Quadratic fluctuations}

The giant magnon solution of \cite{HM}, when presented in an
appropriate gauge looks like a localized soliton. The spatial extent
of the solution
is $J$ (or $J+a E$ depending on the choice of light-cone gauge) and
the original solution of \cite{HM} was defined on the infinite line
i.e. with $J=\infty$. Once we consider finite but large $J$, the
solution will be deformed and the resulting correction to the energy
was found in \cite{AFZ}. If we want to compute the sum of energies of
quadratic fluctuations, in principle we should consider fluctuations
around the finite $J$ solution.

Let us note, however, that according to L{\"u}scher formulas the
leading finite size corrections appear with definite exponential terms
which differ between the $\mu$-term, associated more with deformations
of a classical solution and the $F$-term which appears to have an
exponential term characteristic of quadratic fluctuations around the
spinning string. As in this paper we want to concentrate on the term
with the latter characteristics, we will neglect the impact of the
deformation of the solution at finite $J$ on the energies of
fluctuation modes. Such an effect would have generically an
exponential factor of the $\mu$-term type.

We will hence consider fluctuations around the infinite $J$ magnon and put
them on a cylinder with periodic boundary conditions of circumference $J$.
For general theories there
is also a clear distinction between $\mu$ and $F$-term exponential
scalings so we will proceed with this assumption.

The setup is in fact very similar to the one considered in the paper
\cite{Dorey} where effects of fluctuations around the
$J=\infty$ magnon were analyzed. However putting the fluctuations on a
cylinder in
\cite{Dorey}  was only a regularization procedure prior to the limit
$J \to  \infty$. Here we would like to argue that it can also be used to
obtain leading finite-size effects of quantum fluctuations.

Since the soliton is localized, very far from the soliton core the
fluctuation will be just like a fluctutation around the vacuum, hence
in this case another soliton. Now we have to impose the periodicity
condition for the wave function of the `fluctuation soliton'. As it
will pass around the
cylinder it will get an additional phase shift from scattering
with the `giant magnon' which can be directly expressed in terms of
the forward S-matrix:
\eq
\label{e.sdl}
S^{ba}_{ba}(k,p)=e^{i \dl_{ba}(k,p)}
\eqx
where the original soliton (`giant magnon') is of type `a', while the
fluctuation is an excitation of type `b'. We assume that $e^{i
\dl_{ba}(k,p)}$ is a pure phase, which is true for diagonal scattering and for small fluctuations around the giant magnon  (see Eqns 115-117 in \cite{Dorey}). The quantization condition
then reads
\eq
\label{e.quantization}
k_{n}=\frac{2\pi n}{L}+\frac{\delta_b(k_{n})}{L}
\eqx
where we denote the circumference by $L$ and we suppress the explicit
$p$ and $a$ dependence of $\dl_{ba}(k,p)$. The scattering phase is taken at $L = \infty$ because the corrections to $\dl_{ba}$ vanishing at infinity will be subleading to our result.
A summation over the
zero-point energies would then be
\eq
\label{e.naive}
\dl \eps_{naive} =\f{1}{2} \sum_b \sum_{n=-\infty}^\infty (-1)^{F_b}
\left(\eps(k_{n}) -\eps(k_{n}^{(0)})\right)
\eqx
where we subtracted off fluctuations around the vacuum with the
standard momenta
\eq
k_{n}^{0}=\frac{2\pi n}{L}
\eqx
We will argue that to have agreement with the $F$-term computation we
still have to slightly modify (\ref{e.naive}).

In the present context, the giant magnon is not stationary but is
moving with velocity
\eq
v=\f{d\eps(p)}{dp}
\eqx
The whole system is periodic if one considers {\em together} time
translations $t \to t+\tau$ and space translation $x \to x+L$, where
the period is $\tau=L/v$. The analogs of the standard phase factor
$\omega \tau \equiv \eps(k)\tau$ are now the stability angles
\cite{Dashen}
\eq
\nu(k)=\tau \eps(k)+kL \equiv \tau \eps(k) +\dl(k)=\tau\left( \eps(k)
+\f{v}{L} \dl(k) \right)
\eqx
where we used the fact that $e^{2\pi i n}=1$. This suggests that the
correct quantity to sum is
\eq
\label{e.final}
\dl \eps_{final}=\f{1}{2} \sum_b \sum_{n=-\infty}^\infty (-1)^{F_b}
\left(\eps(k_{n}) +\f{v}{L} \dl(k_{n}) \right) -(\mbox{vacuum})
\eqx
We will proceed with this assumption and show that the generalized $F$-term
formula (\ref{e.fgen}) is exactly the leading exponential behaviour of
the above sum. In the derivation we will just make a mild assumption
that $\eps(k)$ is a symmetric function of $k$.

\section{Summation over fluctuations}

The main technical obstacle in calculating the sum (\ref{e.final}) is
that in almost all cases we are unable to solve analytically the
quantization conditions for the momenta
(\ref{e.quantization}). However this problem may be bypassed by
writing an iterative but exact solution. Let us denote the combination
$2\pi n/L$ by $t$. Then it is clear that an exact solution of
(\ref{e.quantization}) is
\eq
k(t)=t+\frac{\delta(t+\frac{\delta(t+\ldots)}{L})}{L}
\eqx
Now we can perform a Poisson resummation of the sum over $n$:
\eq
\sum_{n=-\infty}^\infty F\left(\f{2\pi n}{L} \right) = \f{L}{2\pi}
\sum_{m=-\infty}^\infty  \int_{-\infty}^{+\infty} F(t) e^{-im L t} dt
\eqx

\subsection*{L\"{u}scher formulas from fluctuations}

In this subsection we will concentrate on the terms which give the
leading exponential
large $L$ corrections, namely terms with $m=\pm 1$. As a result we
will get the L{\"u}scher's F-term. It turns out that it is possible to
include also subleading terms in a closed form (see the next
subsection).\\
Let us first consider the summation over energies in
(\ref{e.final}). To save space we will reinstate the summation over
types of fluctuations and $(-1)^{F_b}$ at the end of the calculation.
We thus have
\eq
\dl \eps_{1}=\frac{L}{4\pi}\int_{-\infty}^{+\infty}e^{i L t}
(\epsilon(k(t)) -\epsilon(t)) dt+\frac{L}{4\pi}
\int_{-\infty}^{+\infty} e^{-i L t}(\epsilon(k(t))-\epsilon(t)) dt
\eqx
Now after an integration by parts,
we obtain
\eq
-\frac{1}{4\pi i}\int_{-\infty}^{+\infty}e^{i L t} \epsilon'(k(t))
\frac{dk}{dt} dt + \frac{1}{4 \pi i}\int_{-\infty}^{+\infty}e^{-i L t}
\epsilon'(k(t)) \frac{dk}{dt} dt
-\frac{1}{2\pi i}\int_{-\infty}^{+\infty}e^{-iLt}\epsilon'(t) dt
\eqx
The next and key step is to change the integration variables in the first two
integrals from $t$ to $k$ and use the functional equation
$k(t)=t+\dl(k(t))/L$. The result is
\eq
-\frac{1}{4\pi i}\int_{-\infty}^{+\infty}e^{i L
  (k-\frac{\delta(k)}{L})} \epsilon'(k) dk +\frac{1}{4 \pi i}
\int_{-\infty}^{+\infty}e^{-i L (k-\frac{\delta(k)}{L})} \epsilon'(k) dk
-\frac{1}{2\pi i}\int_{-\infty}^{+\infty}e^{iLt}\epsilon'(t)dt
\eqx
Since $\epsilon'(t)$ is antisymmetric this can be
rewritten as
\eq
\label{e.first}
\dl \eps_{1}=\frac{1}{4 \pi i}\int_{-\infty}^{+\infty} e^{-i L k}
(e^{i\delta(k)} +e^{-i\delta(-k)}-2) \epsilon'(k) dk
\eqx
Using the relation between the forward $S$ matrix and the phase shifts
we obtain something very similar to the $F$-term:
\eq
\dl \eps_{1}=\frac{1}{4 \pi i}\int_{-\infty}^{+\infty} e^{-i L k}
(S^{ba}_{ba}(k,p) +(S^{-1})^{ba}_{ba}(-k,p)-2) \epsilon'(k) dk
\eqx
where we use the relation
\eq
(S^{-1})_{ab}^{ab}(k,p)=\f{1}{S_{ab}^{ab}(k,p)}
\eqx
which is valid for diagonal matrices.
We will now show that the second term in the above equation, when
summed over all flavours with $(-1)^{F_b}$ is in fact equal to the first
one.
\eq
\label{e.equality}
\sum_{b}(-1)^{F_{b}}\cdot (S^{-1})_{ba}^{ba}(-k,p)=\sum_{b}(-1)^{F_{b}}S_{ba}^{ba}(k,p)
\eqx
This equality comes from using crossing symmetry of the S-matrix
\eq
\label{a.crossing}
(C^{-1}\otimes 1)S^{st_{1}}(p_{1},p_{2})(C\otimes 1) S(-p_{1},p_{2})=1
\eqx
where $C$ is charge conjugation matrix and $^{st_{1}}$ denotes the
supertranspose in the first entry of S:
\eq
(S^{st_{1}})^{b_{1}b_{2}}_{a_{1}a_{2}}=(-1)^{F_{a_{1}}F_{b_{1}}+F_{a_{1}}}
S^{a_{1}b_{2}}_{b_{1}a_{2}}
\eqx
Using crossing symmetry (\ref{a.crossing}) we can rewrite
\eq
\sum_{b}(-1)^{F_{b}}\cdot(S^{-1})_{ba}^{ba}(-k,p)=
sTr_{1} (C^{-1}S^{st_{1}}(k,p)C)^a_a
\eqx
where $sTr_{1}$ denotes the supertrace with respect to the first entry
of S. Using the properties of supertrace we can change the order of
the matrices to obtain
\eq
sTr_{1}(C^{-1}S^{st_{1}}(k,p)C)^a_a
=sTr_{1}(CC^{-1}S^{st_{1}}(k,p))^a_a
=\sum_{b}(-1)^{F_{b}}S_{ba}^{ba}(k,p)
\eqx
where in the last equality we used the fact that $(-1)^{F_{b}F_{b}+F_{b}}=1$.
Taking into account the above we obtain finally for $\dl \eps_1$:
\eq
\label{e.epsi}
\dl \eps_1=\frac{1}{2 \pi i}\int_{-\infty}^{+\infty} \epsilon'(k) e^{-i L k}
\sum_b (-1)^{F_b} (S^{ba}_{ba}(k,p)-1)  dk
\eqx
In order to transform the above integral into a $F$-term like integral
one has to shift the contour of integration in the same way as in the
derivation of the $F$-term so as to make the momentum to be purely
imaginary. The question of boundary terms is nontrivial and has to be
considered on a case by case basis. However it certainly works for
relativistic theories and, more importantly in the present context, it
also works for a fermion
system with the giant magnon dispersion relation (see
\cite{US}). Incidentally the integrality of $L$ was necessary
there. It would be interesting to study this point further.

Note that at this stage we are missing the second piece of
(\ref{e.fgen2}). We will show now that it comes from performing Poisson
resummation of the second term in (\ref{e.final}).

To this end we have to evaluate
\eq
\dl \eps_2 = \frac{\epsilon'(p)}{4\pi}\int_{-\infty}^{+\infty} e^{i
  L t} \delta(k(t))dt + \frac{\epsilon'(p)}{4\pi}
\int_{-\infty}^{+\infty} e^{-i L t}\delta(k(t))dt
\eqx
Using the quantization condition (\ref{e.quantization}) it is
convenient to express $\dl(k(t))$ as
\eq
\dl(k(t))=k(t)-t
\eqx
Plugging it into the
above integral and performing similar manipulations as for
$\dl \eps_1$, we arrive at
\eq
\label{e.epsii}
\dl \eps_2 = \frac{\epsilon'(p)}{4\pi i}\int_{-\infty}^{+\infty}e^{-i
  L k}(e^{i\delta(k)}+e^{-i\delta(-k)}-2) dk
\eqx
Combining this contribution with (\ref{e.first}), expressing the phase
shifts through the $S$-matrix, rotating contours,
using the equality (\ref{e.equality}) we arrive at the complete
expression for the $F$-term:
\eq
\dl \eps_1+\dl \eps_2=\int_{-\infty}^{\infty}\frac{\mbox{d}\qs}{2 \pi
  i}\left(\eps'(\qs)-
\eps'(p) \right)\cdot e^{-i \qs L }\cdot \sum_b (-1)^{F_b}
\left( S_{ba}^{ba}(\qs,p)-1 \right)
\eqx

The above derivation shows that L{\"u}scher's $F$-term is equivalent to a
summation over fluctuations. In order to recover the full expression
we have to consider energies derived from stability angles. A special
case of the above formula for a relativistic dispersion relation leads
to the formula of \cite{KM} for corrections to a moving particle,
while further specializing to a particle at rest reduces to the
classical L{\"u}scher formula \cite{Luscher}. In the following section we
will apply this formalism to calculate the leading correction to the
giant magnon dispersion relation coming from quadratic fluctuations.

\subsection*{Refinements}

It is straightforward to compute contribution to the energy shift from
the terms with arbitrary value of $m >0$.\footnote{The case $m = 0$ is
trivial. There is no contribution to the energy shift under the
assumption that $\sum_{b}
(-1)^{F_{b}} \dl_{b a}(k) = 0$ for any $a$ and $k$ (this is true for the giant
magnon phase shifts).} As a result we get

\eq
\label{e.m}
\dl \eps^{(m)}=\frac{1}{4 \pi i m}\int_{-\infty}^{+\infty} e^{-i m L
  k}\left( \epsilon'(k)-\epsilon'(p)\right)
(e^{i m \delta(k)} +e^{-i m \delta(-k)}-2)  dk
\eqx
where $m$ has the interpretation of the winding number associated with
the virtual soliton going around the cylinder and interacting
$m$-times with the giant magnon (which is still a 1-loop result). The
L{\"u}scher's $F$-term is reproduced by $m = 1$ (this corresponds to
a single interaction with a virtual particle).

In order to obtain the Poisson-resummed $1$-loop energy shift we have
to sum over all values of $m$. After the identification (\ref{e.sdl})
and using crossing symmetry the final formula reads
\eq
\sum_{m=1}^{+\infty}\dl \eps^{(m)}= - \int_{-\infty}^{\infty}\frac{\mbox{d}\qs}{2 \pi
  i}\left(\eps'(\qs)-
\eps'(p) \right) \sum_b (-1)^{F_b}
 \log\Big( \frac{1-S_{ba}^{ba}(\qs,p)e^{-i \qs L }}{1-e^{-i \qs L }}\Big)
\eqx
We have to keep in mind that terms with higher $m$ will appear at the same order as contributions from higher loop processes which are not taken into account by the sum over fluctuation energies.

Let us now return to the formulas (\ref{e.epsi}) and
(\ref{e.epsii}). These provide integral formulas equivalent to
L\"{u}scher formulas but defined as integrals over {\em physical} real
momenta. Although our derivation and interpretation in terms of
fluctuations fails for general S-matrices with nondiagonal scattering
we found that one can give a similar integral formula which is valid
also in these other cases (a typical example would be e.g. the O(3)
model)
\eq
\dl\eps = \f{1}{2\pi i} \int dk (\eps'(k)-\eps'(p))e^{-ikL}
\sum_b(-1)^{F_b}\left( S_{ba}^{ba}(k,p)-1 \right)
\eqx

\section{The giant magnon}

Let us now proceed to compute the leading finite size correction to
the giant magnon dispersion relation coming from quantum
fluctuations. We will perform a saddle point calculation of the
$F$-term integral. The exponent in the $F$-term formula is
\eq
e^{-2J \arcsinh \left( \f{1}{4g} \sqrt{1+q^2} \right)}
\eqx
where we use the conventions of \cite{BES} i.e. $g=\sqrt{\lm}/(4\pi)$.
Saddle point expansion gives
\eq
e^{-\f{J}{2g} -\f{J}{4g} q^2}
\eqx
The saddle point has Euclidean energy $q=0$. Gaussian integration
gives
\eq
2\sqrt{\f{\pi g}{J}} e^{-\f{J}{2g}} = e^{-\f{2\pi J}{\sqrt{\lm}}} \cdot
\f{\lm^{\f{1}{4}}}{\sqrt{J}}
\eqx
The rest of the integrand has to be evaluated at the saddle point
$q=0$. The Jacobian factor is then trivial
\eq
\left(1-\f{\eps'(p)}{\eps'(\qs)} \right) \to 1
\eqx
The phase shifts have to be evaluated for the virtual particle at the
saddle point:
\eq
x_q^+ = i\left(1+\f{1}{4g}\right) \qqqq
x_q^- = i\left(1-\f{1}{4g}\right)
\eqx
and for the giant magnon described by the strong coupling expressions:
\eq
x_p^+ = e^{\f{ip}{2}} \left(1+\f{1}{4g \sinpt}\right) \qqqq
x_p^- = e^{\f{-ip}{2}}\left(1+\f{1}{4g\sinpt}\right)
\eqx
Now we may evaluate the phase shifts using (\ref{e.sdl}). The
contribution of the dressing phase at the saddle point reduces, in the
strong coupling limit, to the contribution of the AFS phase \cite{AFS}
which gives
\eq
\f{1+\sinpt}{1-\sinpt} e^{-ip} e^{-2\sinpt}
\eqx
The phase shifts of the $S^5$ scalars, $AdS_5$ scalars and fermions (see Eqns 36-41 in \cite{Dorey})
then evaluate to
\eqn
\left(e^{i\dl}\right)_{S^5} &=& \f{1+\sinpt}{1-\sinpt} \cdot e^{-2\sinpt}\\
\left(e^{i\dl}\right)_{AdS_5} &=& 1\cdot e^{-2\sinpt} \\
\left(e^{i\dl}\right)_{fermions} &=& \f{\cos \f{p}{4}+\sin \f{p}{4}}{ \cos
  \f{p}{4}-\sin \f{p}{4}} \cdot e^{-2\sinpt}
\eqnx
Summing the phase shifts $\left(4\left(e^{i\dl}\right)_{S^5} +4
\left(e^{i\dl}\right)_{AdS_5} -8 \left(e^{i\dl}\right)_{fermions}
\right)$ gives the final expression
\eq
\dl \eps^F= -\f{1}{2\pi}\cdot \f{\lm^{\f{1}{4}}}{\sqrt{J}} \cdot \f{16
  \sin^2 \f{p}{4}}{1-\sinpt}  e^{-2\sinpt} e^{-\f{2\pi J}{\sqrt{\lm}}}
\eqx
The above procedure is of course equivalent to performing a saddle
point directly in the Poisson resummed expression for a sum over
fluctuation energies. To this order of approximation it is enough to
consider relativistic dispersion relation for the fluctuations as in
\cite{Dorey}, $\eps(k)=\sqrt{1+k^2}$. Performing a change of
variables to $k=\f{2r}{1-r^2}$ one can evaluate the saddle point to
$r=i$. The derivation in section 3 however, identifies directly the
sum over fluctuations which corresponds to the $F$-term formula in its
full generality without the need for a saddle point approximation.

\section{Discussion}

In this paper we have shown that the generalized L{\"u}scher $F$-term
formula (\ref{e.fgen}) has a very transparent interpretation as the
leading exponential term in a summation over frequencies derived from
stability angles. For a particle at rest, corresponding to the
standard L{\"u}scher calculation, the summation is just over zero-point
energies. For a moving particle, the modifications due to the
stability angles are necessary.
Hence under very mild assumptions such calculations
of 1-loop effects have to agree.

We have used the above formalism to evaluate the leading finite size
correction to the giant magnon dispersion relation coming from quantum
fluctuations. By the above reasoning the F-term computation and the
summation over modes are by definition really identical computations.

It would be interesting to understand more deeply the appearance of
these effective energies derived from stability angles in this
context especially as the relation to WKB methods is not entirely
clear here. We leave this problem for future research.

\bigskip

\noindent{\bf Acknowledgments.} RJ would like to thank Zoltan Bajnok
and Sakura Sch{\"a}fer-Nameki for interesting discussions. This
work has been supported in part by Polish Ministry of Science and
Information Technologies grant 1P03B04029 (2005-2008), RTN network
ENRAGE MRTN-CT-2004-005616, and the Marie Curie ToK KraGeoMP (SPB
189/6.PR UE/2007/7).

\bigskip

\noindent{\bf Note added:}
As this paper was being completed, an interesting paper
\cite{Vieira} appeared were quantum fluctuations
around the giant magnon were calculated and
compared with the $F$-term using different methods.

\end{document}